\documentclass[a4paper,twoside,10pt]{letter}
\usepackage{graphicx,saj,multicol,subeqnarray}

\def\papertype{Original Scientific Paper}

\def\udc{520.822 +524.7 NGC2366 +524.354}
\begin{document}
\baselineskip=3.1truemm
\columnsep=.5truecm
\newenvironment{lefteqnarray}{\arraycolsep=0pt\begin{eqnarray}}
{\end{eqnarray}\protect\aftergroup\ignorespaces}
\newenvironment{lefteqnarray*}{\arraycolsep=0pt\begin{eqnarray*}}
{\end{eqnarray*}\protect\aftergroup\ignorespaces}
\newenvironment{leftsubeqnarray}{\arraycolsep=0pt\begin{subeqnarray}}
{\end{subeqnarray}\protect\aftergroup\ignorespaces}

\makeatletter
\def\blfootnote{\xdef\@thefnmark{}\@footnotetext}
\makeatother


\markboth{\eightrm OPTICAL OBSERVATIONS OF NGC 2366 GALAXY} {\eightrm M. M. VU\v{C}ETI\'{C} et al.}

{\ }

\publ

\type

{\ }


\title{OPTICAL OBSERVATIONS OF THE NEARBY GALAXY NGC 2366 THROUGH NARROWBAND H$\alpha$ AND [S$\mathbf{II}$] FILTERS. SUPERNOVA REMNANTS STATUS$^{*}$}
\blfootnote{$^{*}$Based on data collected with 2-m RCC telescope at Rozhen National Astronomical Observatory}


\authors{M. M. Vu\v{c}eti\'{c}$^{1}$, D. Oni\'{c}$^{1}$, N. Petrov$^{2}$, A. \'{C}iprijanovi\'{c}$^{1}$ and M. Z. Pavlovi\'{c}$^{1}$}

\vskip3mm


\address{$^1$Department of Astronomy, Faculty of Mathematics,
University of Belgrade\break Studentski trg 16, 11000 Belgrade,
Serbia}

\Email{mandjelic}{math.rs, donic@math.rs, aleksandra@math.rs, marko@math.rs}

\address{$^{2}$Institute of Astronomy and National Astronomical Observatory, Bulgarian
Academy of Sciences, 72 Tsarigradsko Shosse Blvd, BG-1784 Sofia, Bulgaria}

\Email{nip.sob}{gmail.com}


\dates{January 31, 2019}{February 22, 2019}


\summary{We present detection of 67 \hbox{H\,{\sc ii}} regions and two optical supernova remnant  (SNR) candidates  in the nearby irregular galaxy NGC 2366.
The SNR candidates were detected by applying \hbox{[S\,{\sc ii}]}/H$\alpha$ ratio criterion to observations made with the 2-m RCC telescope at
Rozhen National Astronomical Observatory in Bulgaria. In this paper we report coordinates, diameters, H$\alpha$ and \hbox{[S\,{\sc ii}]} fluxes
for detected objects across the two fields of view in NGC 2366 galaxy. Using archival XMM-Newton observations we suggest possible X-ray counterparts
of two optical SNR candidates. Also,  we discard classification of two previous radio SNR candidates in this galaxy, since they appear to be background galaxies.}



\begin{multicols}{2}
{


\section{1. INTRODUCTION}

Being explosive events that release large amount  of energy {($\sim10^{51}$ergs)} into the interstellar medium (ISM) enriching it with heavy elements, together with their use as standard candles for cosmological distance determination {(single-degenerate type Ia supernovae)}, supernovae (SNe) are extremely interesting phenomena. As interesting are their remnants, which are precious laboratories for { collisionless} shock physics, particle acceleration and magnetic-field { amplification}. By studying a large sample of supernova remnants (SNRs) in a given galaxy, we can also reveal global properties of SNRs and SNe, their interactions and parameters of the ISM - abundances, temperatures, density. {Search for extragalactic SNRs
has an advantage} of avoiding heavy Galactic absorption and also dealing with objects with known distances. Galactic SNR sample suffers hardly due to large distance uncertainties.

Optical extragalactic searches for SNRs are mainly done by using emission line ratio criterium \hbox{[S\,{\sc ii}]}/H$\alpha>0.4$ {(Mathewson and Clarke 1973, D'Odorico et al. 1980, Fesen et al. 1984,} Matonick and Fesen 1997, Blair and Long 1997). This criterium is valid for shock-heated plasmas - sulfur is found in a wide variety of ionization states in the extended recombination zone behind radiative shock, while in photoionized \hbox{H\,{\sc ii}} regions it is predominantly in S$^{++}$ state. So far, more than 1200 optical SNRs in 25 nearby galaxies up to 10 Mpc have been detected (Vu\v{c}eti\'{c} et al.~2015 and references therein).{ Also, optical observations increased number of the known Galactic SNRs (Stupar et al. 2008, Sabin et al. 2013), which are predominantly discovered in radio-wavelengths. On the other hand, majority of the extragalactic SNRs have been detected in optical wavelengths, as shown on Venn diagrams that
summarize the number of SNRs exhibiting emission in the different domains in Bozzetto et al. (2017).} Nevertheless, only a small number of galaxies have been thoroughly surveyed for SNRs in more than one frequency range, and therefore it is hard to claim the real status of SNRs in those galaxies. {Only  Magellanic Clouds, thanks to their vicinity, have majority of SNRs  detected simultaneously in optical, X-ray and radio-domain (Bozzetto et al. 2017).}  That is why multiwavelength detection of SNRs in other galaxies is essential for the confirmation of SNRs' true nature.

In this paper we present optical photometric search for SNRs in NGC 2366 galaxy.  NGC 2366 is a Magellanic barred irregular galaxy of class IB(s)m (de Vaucouleurs et al.~1991). It is also designated as blue compact dwarf  galaxy (BCDG), which are the least chemically evolved gas-rich star-forming galaxies known in the local Universe (Yin et al. 2011).
BCDGs are undergoing intense bursts of star formation, giving birth to thousands of O stars in a very compact starburst region {(see e.~g.~M\'{e}ndez
et al.~1999 for the case of He 10-12 galaxy)}. NGC 2366 is also known as cometary BCDG,
which are characterized by a high surface brightness star-forming region (the comet's head) at one end of an elongated low surface brightness stellar body (the comet's tail).
The chain of \hbox{H\,{\sc ii}} regions extending over galaxy's body is suggestive of self propagating star formation which stopped at the edge of the galaxy. {Finally,
Micheva et al.~(2017) have found that NGC 2366 is an excellent analog of the so called Green Pea galaxies, which are characterized by extremely high ionization parameters.}

\vskip.5cm \noindent
\begin{minipage}{\textwidth}
\parbox{\columnwidth}{

{\bf Table 1.} Properties of NGC 2366 galaxy, \\ taken from NED$^{1}$. \vskip.25cm
\begin{tabular}{@{\extracolsep{0.0mm}}l r @{}}
\hline
 Right ascension (J2000) &  07h28m54.66s \\
 Declination (J2000) & +69$^{\circ}$12$'$68\uu8 \\
 Redshift & 0.00027 \\
 Velocity & 80 km s$^{-1}$\\
 Distance$^{2}$ & 3.44 Mpc \\
 Angular size &  $9.0 '\times 3.5'$ \\
 Magnitude & 10.4 mag\\
 Gal. extinction$^{3}$  & 0.132 mag (B)\\
\hline
\hline
\end{tabular}}
$^{1}${\footnotesize \texttt{http://ned.ipac.caltech.edu/}}\\
$^{2}${\footnotesize Tolstoy et al.~(1995)}\\
$^{3}${\footnotesize Schlafly and  Finkbeiner (2011)} \\

\vskip 7mm
\end{minipage}

Being a place of intense star formation, this nearby galaxy is a good candidate for SNR searches. So far, this galaxy has been surveyed for radio-SNRs. Using the Very Large Array (VLA), Chomiuk and Wilcots (2009) produced maps at 20, 6, and 3.6 cm with synthesized beams of 3\uu7$\times$3\uu7 ($\sim60\times60$ pc), and sensitivity of 20 $\mu$Jy/beam, intended to identify SNRs. {Chomiuk and Wilcots (2009)} considered discrete radio sources with non-thermal spectral index $\alpha\leq-0.2$ (defined as $S_{\nu}\propto \nu^{\alpha}$), which have corresponding H$\alpha$ emission,  as SNRs. Non-thermal radio-sources without optical counterparts they classified as probable distant background radio galaxies, whose redshifted optical emission, they thought, would fall outside narrowband H$\alpha$ filter. {We underline here that there are numerous examples of well known and bright radio SNRs with no detected optical emission, such as HFPK334 in  Small Magelanic Cloud (Crawford et al. 2014), SNR J0528-6714 in Large Magellanic Cloud (Crawford et al. 2010) or like Vela Jr. (Maxted at al. 2018).}  Using criteria mentioned above, Chomiuk and Wilcots (2009) suggested five sources -- {N2366-07, N2366-12, N2366-15, N2366-16 and N2366-18} as SNRs in NGC 2366 galaxy.

In tne next section, we present our observations of NGC 2366 galaxy, intended to confirm radio SNRs, and possibly detect new optical SNR candidates. Furthermore, we used archival
\textit{XMM-Newton} observations of this galaxy to search for possible X-ray counterparts of SNR candidates. We will discuss and comment individual sources of specific interest in Section 3, and summarize our results in Section 4.

\section{2. OPTICAL OBSERVATIONS}

The observations were carried out during multiple observational runs (February 2015, November 2016, March 2017), with the 2-m Ritchey-Chr\'{e}tien-Coud\'{e} (RCC) telescope
at the National Astronomical Observatory (NAO) Rozhen, Bulgaria ($\varphi = 41^\circ 41' 35'' ,\ \lambda = 24^\circ  44' 30'' ,\ h = 1759$ m). The telescope was equipped with
VersArray: 1300B CCD camera with 1340$\times$1300 px array, with plate scale of 0\uu258/px, giving the field of view $5'45''\times 5'35''$.

We observed two fields of view (FOV), in order to cover full extent NGC 2366 galaxy. Centers of the fields of view were: FOV1:  R.A.(J2000) = 07:28:34, Decl.(J2000) = +69:10:35;
FOV2: R.A.(J2000) = 07:28:58, Decl.(J2000) = +69:14:26.

Additionally, we obtained B, V and R images of NGC 2366 galaxy on March 29, 2017, from Astronomical station Vidojevica (ASV). Observations were carried out with 1.4-m Milankovi\'{c}
telescope. We took sets of three 5 min exposures through each filter, with seeing 1\uu7 -- 2\uu0.  Composite image obtained at ASV, with marked positions of FOV1 and FOV2 taken from NAO Rozhen,
is given in Fig. 1. Size of the FOV of Milankovi\'{c} telescope, with Apogee U42 CCD attached, is $9'\times 9'$.

The observations from NAO Rozhen were performed with the narrowband \hbox{[S\,{\sc ii}]}, H$\alpha$ and red continuum filters, each wide approximately 30 \AA. FOV1 was observed with
total exposure times of 160 min (continuum), 100 min (H$\alpha$) and 60 min (\hbox{[S\,{\sc ii}]} filter), with median seeing of 2\uu5 -- 3\uu25. FOV2 was observed with total
exposure times of 125 min (continuum), 135 min (H$\alpha$) and 60 min (\hbox{[S\,{\sc ii}]} filter), with median seeing of 1\uu75 -- 2\uu0. Images of standard star Feige 34,
as well as sky flat-field images, were also taken.

}
\end{multicols}

\centerline{\includegraphics[bb=  0 0 282 274, width=0.75\textwidth,
keepaspectratio]{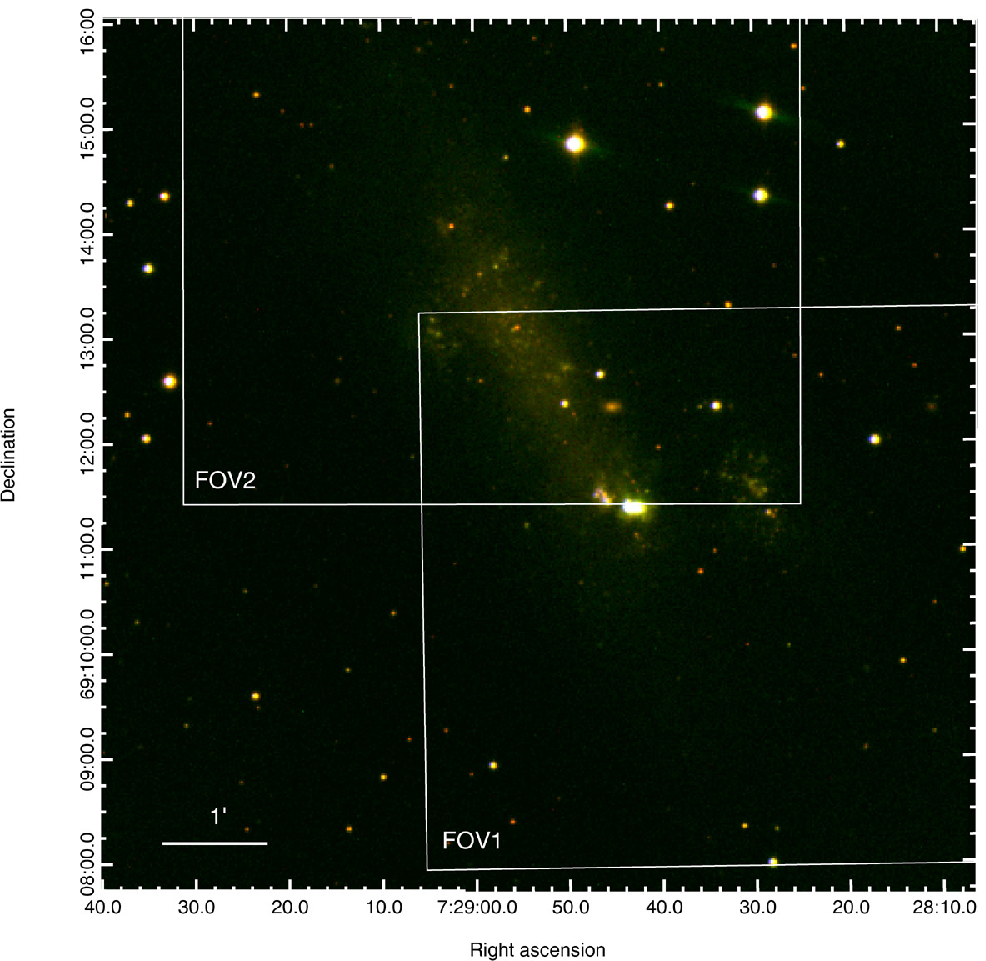}}

\figurecaption{1.}{Composite image of NGC 2366 galaxy, taken with Milankovi\'{c} telescope. Red color is for R filter, green is for V filter and blue is for B filter. Size of the FOV is $9'\times 9'$. FOV1 and FOV2 were observed from NAO Rozhen.}

\vskip 7mm

\begin{multicols}{2}
{
Data reduction was done using standard procedures in IRAF\footnote{IRAF is distributed by the National Optical Astronomy Observatory, which is operated by the Association of Universities
for Research in Astronomy, Inc., under cooperative agreement with the National Science Foundation.} and IRIS\footnote{Available from {\footnotesize \texttt{http://www.astrosurf.com/buil/}}}.
Images through each filter were firstly combined using sigma-clipping method, then sky-subtracted, and finally flux calibrated using the observations of the standard star Feige 34 and its
fluxes from Oke (1990). Afterwards, the continuum contribution was removed from the H$\alpha$ and \hbox{[S\,{\sc ii}]} images, scaling each image to have the same flux for 20 foreground stars in the field, prior to continuum subtraction. H$\alpha$ images were corrected for the filter transmission and  contamination by \hbox{[N\,{\sc ii}]} lines at $\lambda$6548 \AA, $\lambda$6583 \AA, using median \hbox{[N\,{\sc ii}]}$\lambda$6548/H$\alpha$ ratio obtained from spectra of 61 \hbox{H\,{\sc ii}} regions detected in NGC 2366 by Roy et al. (1996). Since there were no previous optical SNR detections, we used same \hbox{[N\,{\sc ii}]}$\lambda$6548/H$\alpha$ ratio for our SNR candidates. Also, emission from \hbox{[S\,{\sc ii}]} lines at $\lambda$6717 \AA, $\lambda$6731 \AA, was corrected for filter transmission, as suggested in Vu\v{c}eti\'{c} et al.~(2013).

\subsection{2.1 H$\alpha$ detected objects}

In Figs. 2 and 3 we present continuum subtracted H$\alpha$ images of FOV1 and FOV2, {and in Fig. 4 we present continuum subtracted \hbox{[S\,{\sc ii}]} image of FOV2, which shows all sources with \hbox{[S\,{\sc ii}]} emission.} Displayed images are zoomed-in, in order to show only regions where  emission nebulae were detected. We detected 67 H$\alpha$ sources, most probably \hbox{H\,{\sc ii}} regions, filaments and superbubbles.

}
\end{multicols}

\centerline{\includegraphics[bb=  0 0 305 283, width=0.85\textwidth,
keepaspectratio]{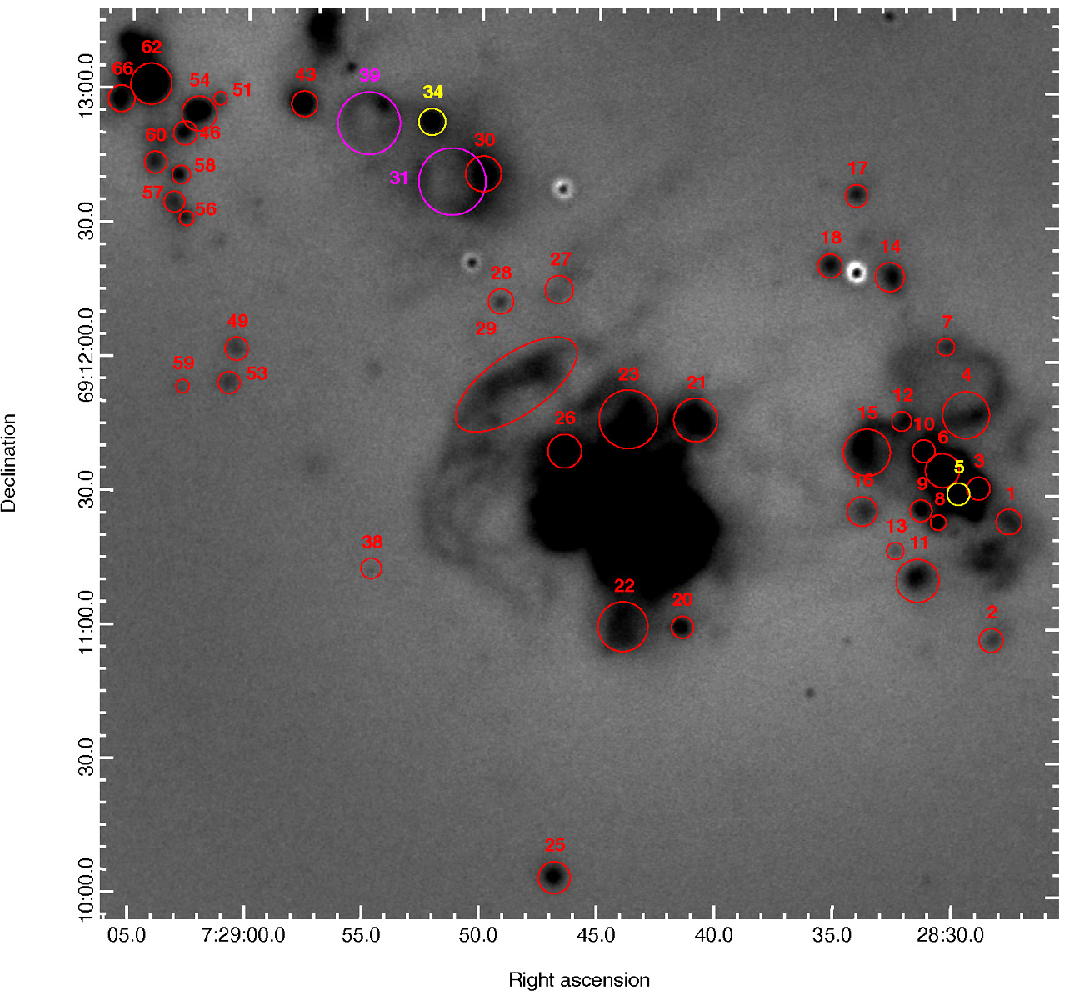}}

\figurecaption{2.}{H$\alpha$-continuum subtracted image of NGC 2366 - FOV1, taken from NAO Rozhen. Properties of marked objects are given in Table 2. {Objects marked with yellow color are SNR {candidates,  while objects marked with magenta color are superbubbles}}. {Non-marked objects are stars and the center of the galaxy, where  continuum was not subtracted well.} }

\begin{multicols}{2}
{
We suggest two sources (Vu19 5 and Vu19 34) as optical SNR {candidates}, according to their enhanced \hbox{[S\,{\sc ii}]} emission, {and detected counterparts in radio and X-ray domain (see Section 3)}. Coordinates, diameters, H$\alpha$ fluxes and \hbox{[S\,{\sc ii}]}/H$\alpha$ ratios for detected objects are given in Table 2. We emphasize here that \hbox{[S\,{\sc ii}]}/H$\alpha$ ratios for SNR candidates are probably underestimated, since H$\alpha$ fluxes of SNR candidates given in Table 2 are corrected for  \hbox{[N\,{\sc ii}]} contamination   in the same way as \hbox{H\,{\sc ii}} regions, while it is expected that shock-heated objects  have  enhanced \hbox{[N\,{\sc ii}]} lines. This would lead to overestimation of H$\alpha$ flux for SNRs, and therefore underestimation of \hbox{[S\,{\sc ii}]}/H$\alpha$ ratio. {We add that given diameters in Table 2 are  overestimated, since observations were made under seeing conditions of 3\uu25 for FOV1 and 2\uu0 for FOV2, which equals to linear size of 55 pc and 35 pc, respectively.}  Majority of the detected \hbox{H\,{\sc ii}} regions and filaments were already detected by Hodge and Kennicutt (1983), Roy et al. (1996) and Eymeren et al. (2007), but without previously reported H$\alpha$ fluxes. {We propose two objects as superbubbles (Vu19 31 and Vu19 39), according to their diameters  ({240 pc and 250 pc}) and present \hbox{[S\,{\sc ii}]} emission, and one object as giant supershell (Vu19 67), which has size above 500 pc.}

}
\end{multicols}

\centerline{\includegraphics[bb=0 0  255 263, width=0.85\textwidth,
keepaspectratio]{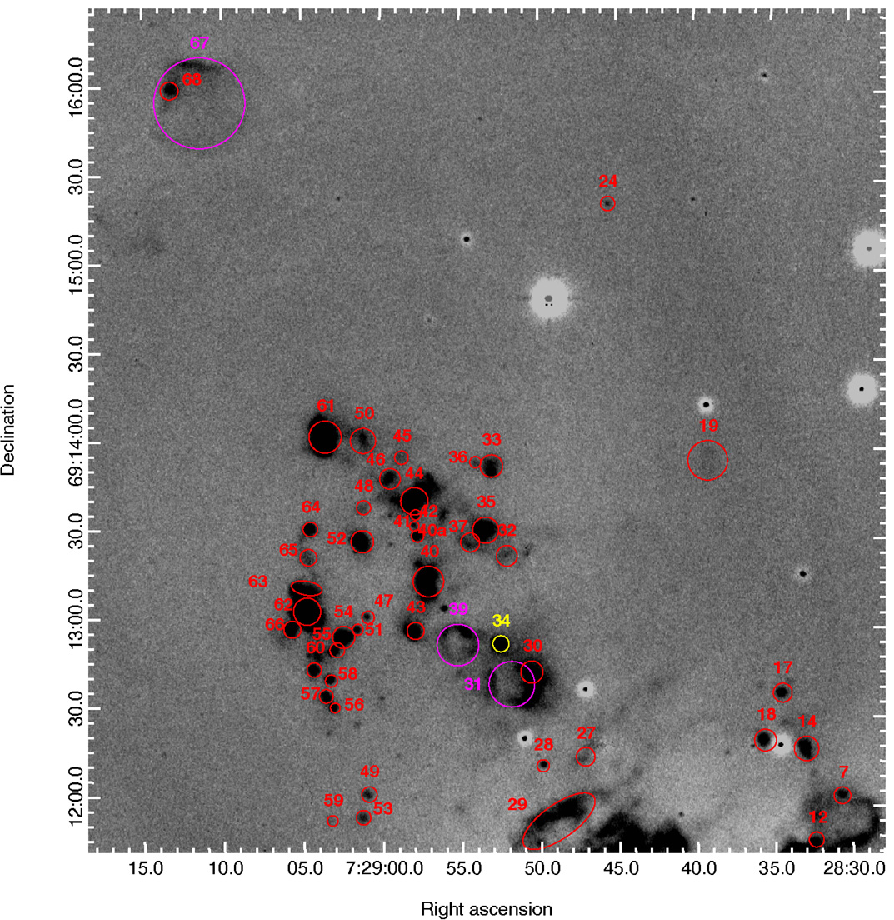}}

\figurecaption{3.}{H$\alpha$-continuum subtracted image of NGC 2366 - FOV2, taken from NAO Rozhen. Properties of marked objects are given in Table 2. {Object marked with yellow color is SNR {candidate, while objects marked with magenta color are superbubbles}. Non-marked objects are stars, where  continuum was not subtracted well.}}

\centerline{\includegraphics[bb= 0 0 299 298, width=0.8\textwidth,
keepaspectratio]{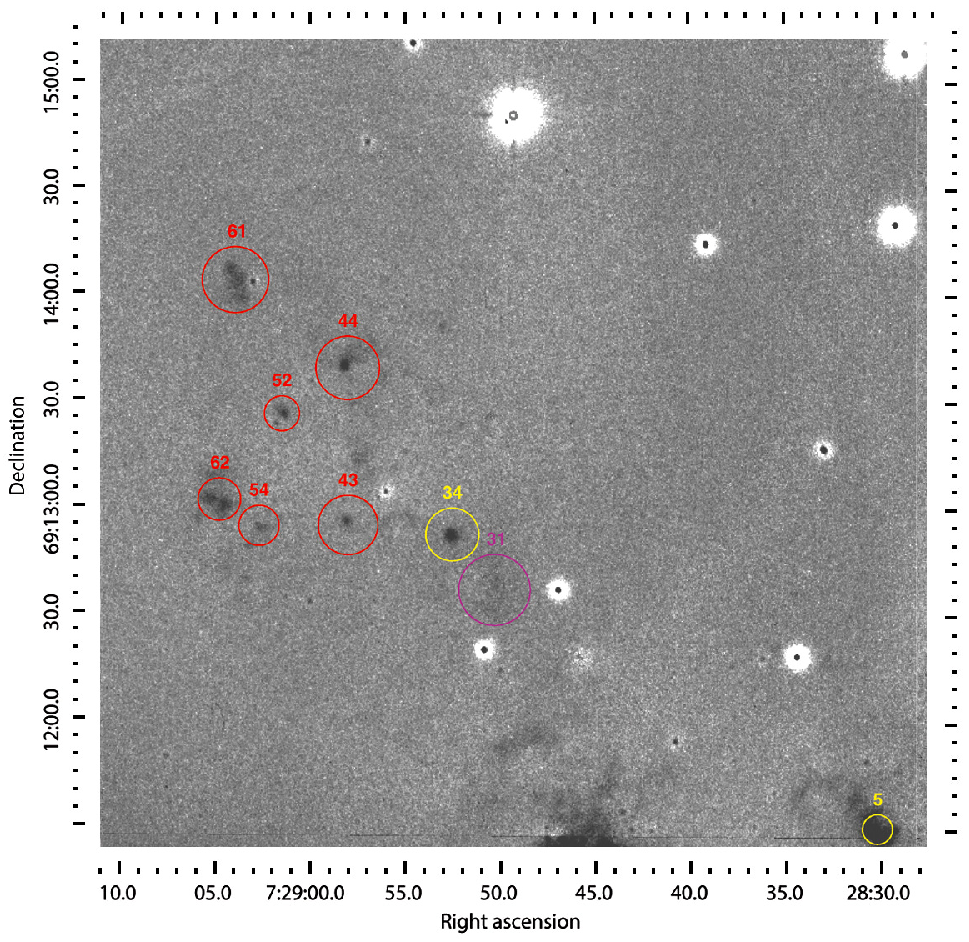}}

\figurecaption{4.}{\hbox{[S\,{\sc ii}]}-continuum subtracted image of NGC 2366 - FOV2, taken from NAO Rozhen. Properties of marked objects are given in Table 2.}

\begin{multicols}{2}
{

Five of our objects (Vu19 5, 27, 34, 38, 44) were detected in radio wavelengths and were suggested to be SNR candidates, based on their radio-spectral index (Chomiuk and Wicots 2009). Our observations also suggest objects Vu19 5 and Vu19 34 as SNR {candidates}, while we discard objects Vu19 27 and Vu19 38 as SNRs (see Section 3 for more details). {In the vicinity of Vu19 27}, which is really marginally detected in H$\alpha$ image, is a bright continuum source, separated about 5 arcseconds from Vu19 27, whose appearance is galaxy-like. This continuum source could match radio source N2366-12 from Chomiuk and Wicots (2009), and since it does not have H$\alpha$ counterpart, it could not be an SNR. This object has also been catalogued as a galaxy  in 2MASS catalogue as 2MASX J07284539+6912186 and a background galaxy N2366BG7 in Drissen et al. (2000). Object Vu19 38 appears only in the continuum image, and is most probably a background galaxy (see Section 3.3 for more details).

\section{3. ARCHIVAL XMM-NEWTON OBSERVATIONS OF NGC 2366}

As previously mentioned, BCDGs are undergoing intense bursts of star formation. Because of the presence of many massive short-lived stars, BCDGs
are expected to emit in the X-rays. This X-ray emission can originate from compact sources such as high-mass X-ray binaries (HMXBs) and/or hot O and Wolf-Rayet stars, or from diffuse sources like hot plasma associated with SNRs {or superbubbles (Sano et al. 2017, Kavanagh et al. 2019)}. NGC 2366 was observed by the {\it XMM-Newton} Observatory with all EPIC cameras using the
medium filter in full field mode (ObsId 0141150201, PI: T.~X.~Thuan).  Thuan et al.~(2014) suggested that NGC 2366 contains two faint X-ray point sources and two faint extended sources. According to them, one point source (XMMU J072858.2+691134) is likely a background AGN, while the other (XMMU J072855.4+691305) {appears to be coincident with a very luminous star associated with NGC 2366. On the other hand, the X-ray luminosity of this object would be comparable to Galactic accreting or colliding-wind HMXBs, suggesting the foreground star scenario. Therefore, identification of object XMMU J072855.4+691305, whose optical counterpart is visible on Fig. 3 as point-source above object Vu19 39, is very intriguing, and still not unique.} The two faint extended sources are  associated with massive \mbox{H\,{\sc ii}} complexes. In Fig. 5 we present an adaptively-smoothed exposure-corrected combined {\it XMM-Newton} EPIC image of the NGC 2366 main parts (scaled with minmax/log and 25 smoothing counts). The image depicts emission detected over the energy range from 0.4 keV to 7.0 keV. Some of the optically detected
sources are represented by white circles. The X-ray data were analyzed using standard tools in the HEASOFT Software Package and the Science Analysis Software (SAS) software package (Version 17.0.0). The SAS tools \verb"epchain" and \verb"emchain" were used to apply standard processing tools to the EPIC datasets, while the tools \verb"mos-filter" and \verb"pn-filter" were used to filter the data for background flaring activity. The effective exposure times of the MOS1/2 and PN cameras were 39 and 29 ks, respectively. {\it XMM-Newton} astrometric accuracy is $\approx$2 -- 4 arcsec (Watson et al. 2009).
}
\end{multicols}

\vskip 3mm

\centerline{\includegraphics[bb= 0 0 1111 628, width=0.8\textwidth,
keepaspectratio]{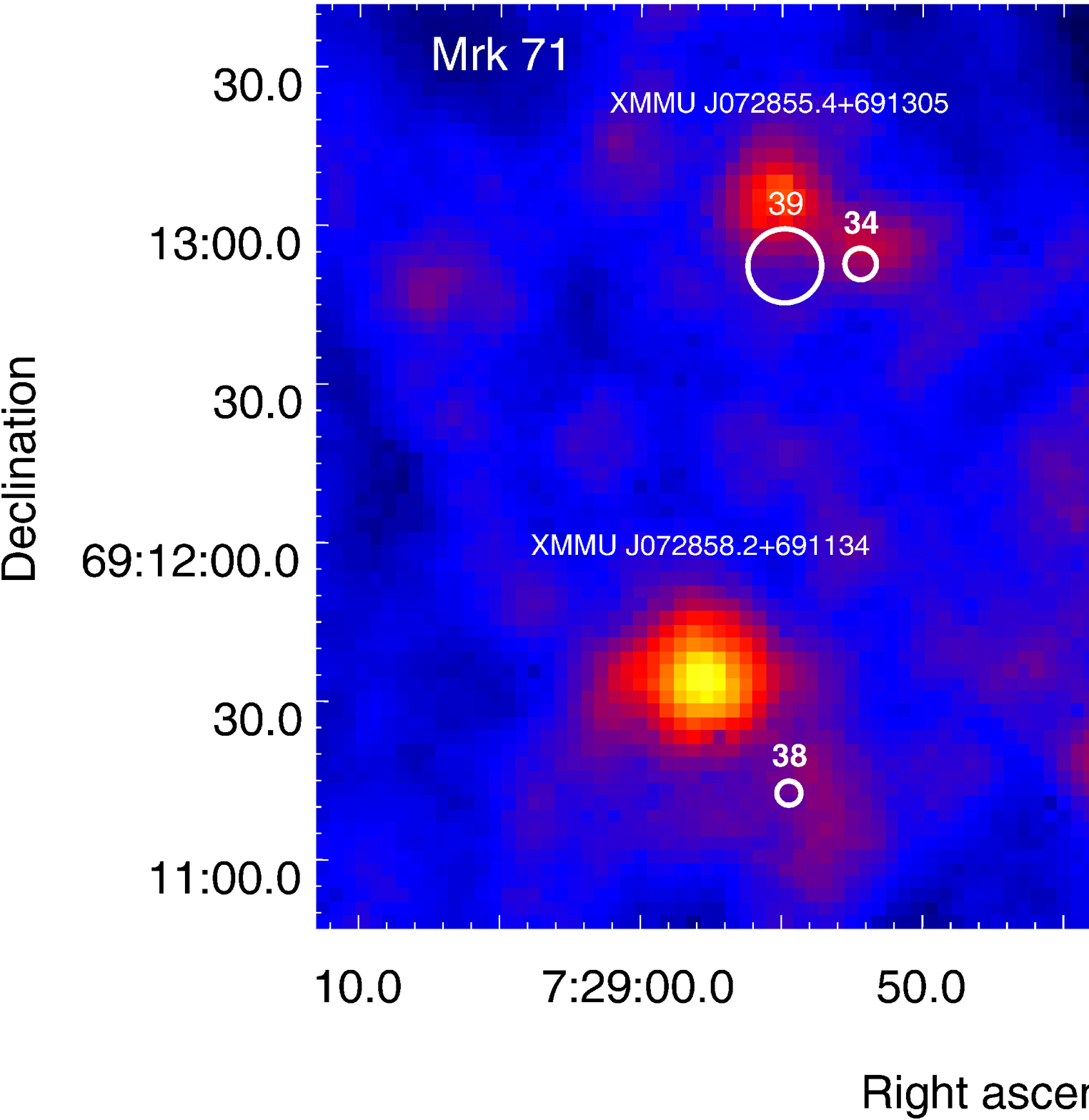}}

\figurecaption{5.}{Adaptively-smoothed exposure-corrected combined {\it XMM-Newton} EPIC image of the NGC 2366 main parts. The image depicts emission detected over the 0.4 keV to 7.0 keV energy range. Some of the optically detected sources are represented by white circles.}

\begin{multicols}{2}
{

\subsection{3.1 SNR candidate Vu19 5}

We have found that our objects Vu19 5 and Vu19  6 are in fact coincident with faint soft X-ray source XMMU J072830.4+691132 (see Fig. 5). Actually, in the analysis of X-ray sources
spatially coincident with NGC 2366, Thuan et al.~(2014) only noted that the extent of this X-ray source aligns well with a dense stellar cluster and a massive \mbox{H\,{\sc ii}} complex,
according to position alignment with the Hubble Space Telescope ({\it HST})  F814W image {(see Fig. 4 in Thuan et al. 2014)}.{ We used \textit{HST} Wide-Field Planetary Camera 2 (WFPC2) image through F656N filter (Proposal ID 6096, PI Drissen L.), to resolve this complex starforming region (also known as NGC2363), where we label our objects 3, 5, 6, 8, 9, and 10. On \textit{HST} WFPC2 image we overlaid \hbox{[S\,{\sc ii}]} contours (Fig. 6). It looks like the position of object Vu19 5 covers at least two H$\alpha$ objects, resolved with \textit{HST}. One is more compact, on the border with object Vu19 3, and the other, on the border with object Vu19 6, exhibits partial-shell structure, with diameter of 62 pc. This shell-like structure is coincident with the maximum of \hbox{[S\,{\sc ii}]} emission, and we propose that this object  could be  the optical SNR {candidate} Vu 19 5, which has an X-ray counterpart. Although measured \hbox{[S\,{\sc ii}]}/H$\alpha$ ratio of this object is  0.21, we emphasis that given \hbox{[S\,{\sc ii}]}/H$\alpha$ ratios are likely underestimated for SNRs (as explained in Section 2.1), and that unresolved area of SNR {candidate} Vu19 5 probably covers  emission of an \mbox{H\,{\sc ii}}, as well. } Unfortunately, the current {\it XMM-Newton} data are not sufficient to definitely characterize the spectral nature of this X-ray emission.
In addition, this source is also included in the {\it XMM-Newton} Serendipitous Source Catalog 3XMM DR8 Version (3XMM J072830.1+691132), {with rather low detection maximum likelihood of 7.45}, and hardness ratios $\mathrm{HR_{1}}>0$ and $\mathrm{HR_{2}}<0$, that are in a good accordance with the conclusions stated in Sasaki et al.~(2018) for known SNRs in the
northern disc of M31 galaxy (see also Sturm et al.~2013 for definition of HRs). {At the position of the border of regions 5 and 6, coincident with the shell,  Chomiuk and Wilcots (2009) detected object N2366-07, characterized as SNR. This object has radio spectral index $\alpha=-0.53\pm0.04$, size of $73\times63$ pc, and flux density 0.20 mJy at 20 cm.}

\subsection{3.2 SNR candidate Vu19 34}

Thuan et al.~(2014) also mentioned faint potential X-ray point source, slightly blended with XMMU J072855.4+691305. This could be an X-ray counterpart of the possible optical SNR, labeled as Vu19 34 in our analysis (see Fig. 5), also detected at radio frequencies (Chomiuk and Wilcots 2009). {Radio source N2366-15, coincident with SNR {candidate} Vu19 34, has spectral index $\alpha=-0.37\pm0.07$, size of $70\times63$ pc, and flux density 0.19 mJy at 20 cm. Optical diameter of object Vu19 34 is 80 pc, which suggest that this SNR {candidate} is in late radiative phase.} With present X-ray data it is not possible to thoroughly examine nature of this X-source. {However, we note that radio, optical and X-ray detection suggests an SNR origin of the source (see Filipovi\'{c} et al. 1998, Long 2017, Bozzetto et al. 2017)}.

\subsection{3.3 Object Vu19 38}

Near the bright X-ray source XMMU J072858.2+691134, likely a background galaxy hosting an AGN, a very faint X-ray emission is detected. Due to the angular proximity,
we cautiously note that this could be an X-ray counterpart to object Vu19 38 (see Fig. 5). This object is most probably a background galaxy, or a pair of interacting galaxies
labeled as N2366BG13 and N2366BG14 in Drissen et al.~(2000), or a compact star cluster (marked as cluster 2 in Billet et al.~2000). Of course, such a conclusion can not
be strongly verified with this particular X-ray observation. Chomiuk and Wilcots (2009) detected radio object N2366-16 at the position of object Vu19 38, and suggested that it is an SNR.
According to its optical appearance, being bright in the continuum filter (at $\lambda$6417 \AA), and its non-detection in H$\alpha$ filter, we definitely discard it as an SNR candidate.

}
\end{multicols}

\centerline{\includegraphics[bb=0 0 503 276, width=1.0\textwidth,
keepaspectratio]{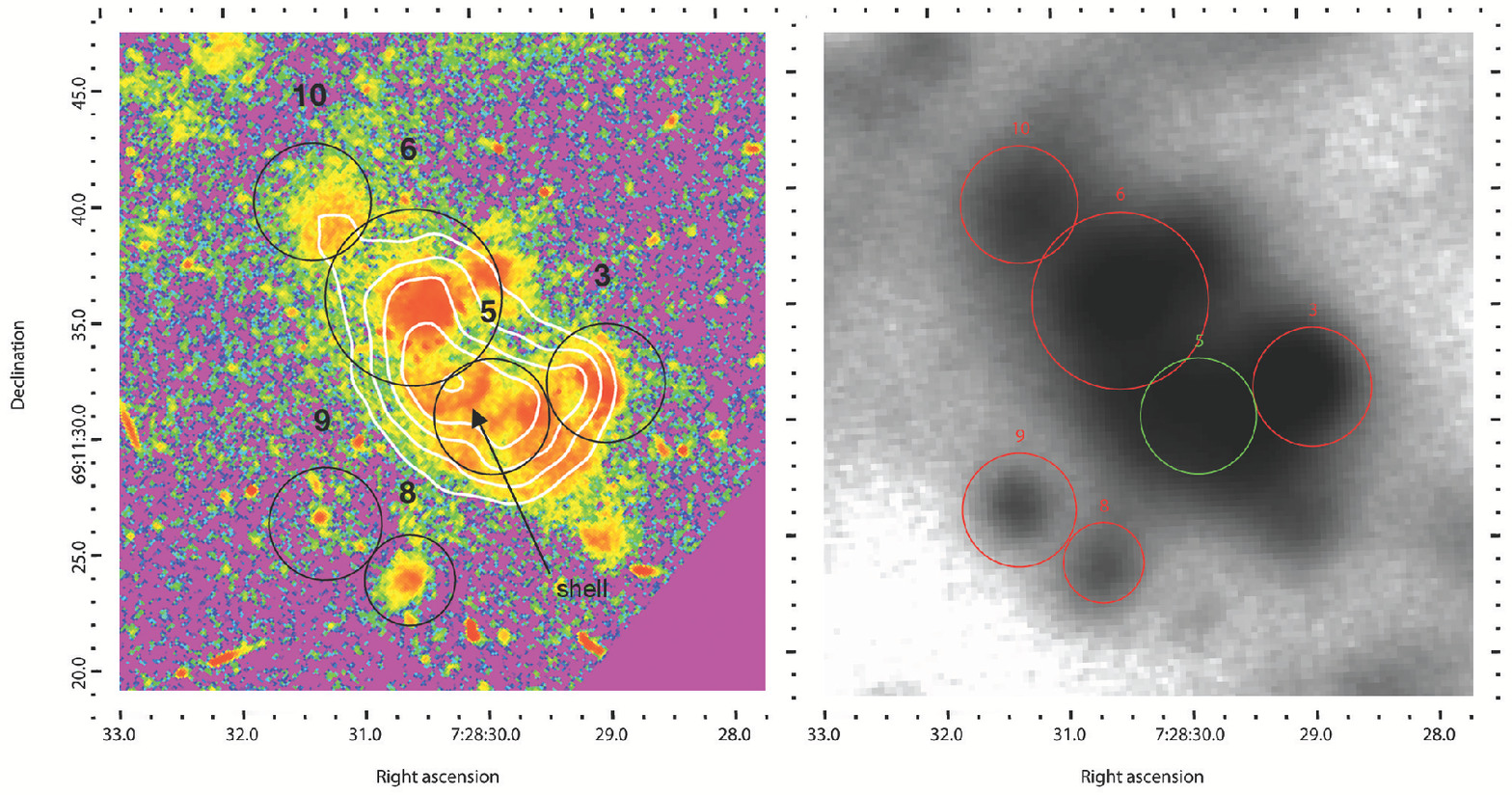}}

\figurecaption{6.}{Left: \textit{HST} WFPC2 F658N image with overlaid \hbox{[S\,{\sc ii}]} contours, centered on the starforming region near object Vu19 5. Maximum of \hbox{[S\,{\sc ii}]}emission aligns with partial  H$\alpha$ shell resolved with \textit{HST}. Right: the same part of our H$\alpha$ image. Seeing was 3\uu25. Diameter of SNR candidate Vu19 5 is 5\uu0=85 pc.}

\noindent
\parbox{\textwidth}{
{\bf Table 2.} Properties of H$\alpha$ emitting regions in NGC 2366 galaxy.
\vskip.25cm
\begin{tabular}{@{\extracolsep{-1.0mm}}l c c c c c l @{}}
\hline
Object & Right   & Decl. & H$\alpha$ flux$^{*}$ & Diameter$^{**}$ & \hbox{[S\,{\sc ii}]}/H$\alpha$  &   Comment \\
ID & Ascension &  &  [erg s$^{-1}$ cm$^{-2}$] &  [pc] & ratio &    \\
& $\alpha _{\mathrm{J2000}}$ & $\delta _{\mathrm{J2000}}$ & $\times10^{-15}$ & &  & \\
\hline \hline
Vu19 1	&	07:28:27.7	&	+69:11:24	&	31.70	&	110	&		&	\mbox{H\,{\sc ii}} region	\\
Vu19 2	&	07:28:28.4	&	+69:10:58	&	1.74	&	90	&		&	\mbox{H\,{\sc ii}} region	\\
Vu19 3	&	07:28:29.0 &		+69:11:32	&	14.61		&	80	& & \mbox{H\,{\sc ii}} region \\
Vu19 4	&	07:28:29.5	&	+69:11:48	&	0.00	&	180	&		&	filament	\\
Vu19 5$^{a}$	&	07:28:30.7	&	+69:11:33	&	11.31	&	62$^{\dag}$	&	0.21	& SNR candidate	\\
Vu19 6	&	07:28:30.5	&	+69:11:36	&	16.70		&	130	& & \mbox{H\,{\sc ii}} region \\
Vu19 7	&	07:28:30.3	&	+69:12:03	&	1.04	&	60	&		&	part of a filament	\\
Vu19 8	&	07:28:30.8	&	+69:11:24	&	2.58	&	60	&		& \mbox{H\,{\sc ii}} region 	\\
Vu19 9	&	07:28:31.4	&	+69:11:27	&	2.73	&	80	&		&	\mbox{H\,{\sc ii}} region	\\
Vu19 10	&	07:28:31.3	&	+69:11:40	&	5.04	&	80	&        & \mbox{H\,{\sc ii}} region \\
Vu19 11	&	07:28:31.5	&	+69:11:11	&	5.91	&	160	&		&	\mbox{H\,{\sc ii}} region	\\
Vu19 12	&	07:28:32.2	&	+69:11:47	&	1.49	&	70	&		& \mbox{H\,{\sc ii}} region		\\
Vu19 13	&	07:28:32.6	&	+69:11:18	&	0.51	&	50	&		&	\mbox{H\,{\sc ii}} region	\\
Vu19 14	&	07:28:32.8	&	+69:12:19	&	4.03	&	70	&		& 	\mbox{H\,{\sc ii}} region	\\
Vu19 15	&	07:28:33.7	&	+69:11:40	&	5.41	&	180	&		&  \mbox{H\,{\sc ii}} reg.	surrounded by filaments	\\
Vu19 16	&	07:28:33.9	&	+69:11:26	&	2.45	&	110	&		& giant \mbox{H\,{\sc ii}} region		\\
Vu19 17	&	07:28:34.2	&	+69:12:37	&	2.32	&	110	&		& giant \mbox{H\,{\sc ii}} 	region	\\
Vu19 18	&	07:28:35.4	&	+69:12:22	&	3.59	&	70	&		& \mbox{H\,{\sc ii}} 	region	\\
Vu19 19	&	07:28:38.9	&	+69:13:55	&	/	&	220	&		&	very faint \mbox{H\,{\sc ii}} region	\\
Vu19 20	&	07:28:41.6	&	+69:11:00	&	3.70	&	80	&		&	\mbox{H\,{\sc ii}} region	\\
Vu19 21	&	07:28:40.9	&	+69:11:46	&	12.72	&	160	&		& giant	\mbox{H\,{\sc ii}} region	\\
Vu19 22	&	07:28:44.0	&	+69:11:00	&	11.05	&	190	&		&	diffuse	\\
Vu19 23	&	07:28:43.8	&	+69:11:47	&	21.32	&	220	&		& giant	\mbox{H\,{\sc ii}} region 	\\
Vu19 24	&	07:28:45.3	&	+69:15:22	&	0.50	&	80	&		&	\mbox{H\,{\sc ii}} region	\\
Vu19 25	&	07:28:46.9	&	+69:10:05	&	4.93	&	120	&		& giant	\mbox{H\,{\sc ii}} region	\\
Vu19 26	&	07:28:46.5 &	+69:11:40 &	3.78		&	130 & & giant \mbox{H\,{\sc ii}} region \\
Vu19 27$^{b}$	&	07:28:46.8	&	+69:12:16	&	/	&	/	&		&	not SNR; see Section 2.1	 \\
Vu19 28	&	07:28:49.2	&	+69:12:13 &		0.61		&	90 & & \mbox{H\,{\sc ii}} region  \\
Vu19 29	&	07:28:48.6	&	+69:11:54	&	0.00	&	$540\times230$	&		&	filament	\\
Vu19 30	&	07:28:50.2	&	+69:11:39	&	8.06	&	230	&		&	giant	\mbox{H\,{\sc ii}} region \\
Vu19 31	&	07:28:51.3 &		+69:12:40	&	5.52	&		250 & 0.11 & superbubble \\
Vu19 32	&	07:28:51.8	&	+69:13:23	&	1.19	&	120	&		&	giant diffuse	\\
Vu19 33	&	07:28:52.7	&	+69:13:53	&	5.54	&	130	&		&	giant \mbox{H\,{\sc ii}} region	\\
Vu19 34$^{c}$	&	07:28:52.3	&	+69:12:53	&	5.47	&	80	&	0.38	&	SNR candidate	\\
Vu19 35	&	07:28:53.1	&	+69:13:31	&	11.16	&	150	&		& giant \mbox{H\,{\sc ii}} region		\\
Vu19 36	&	07:28:53.7	&	+69:13:54	&	0.78	&	60	&		&	\mbox{H\,{\sc ii}} region	\\
Vu19 37	&	07:28:54.1	&	+69:13:27	&	2.02	&	110	&		&	diffuse	\\
Vu19 38$^{d}$	&	07:28:54.7	&	+69:11:12	&	0.15	&	80	&		&	not SNR; see Section 3.3	\\
Vu19 39	& 07:28:54.9 &	+69:12:52 &	5.52	&		240 & 0.09 & superbubble \\
Vu19 40	&	07:28:56.7	&	+69:13:13	& 5.31	&	170	&		&		\mbox{H\,{\sc ii}} region \\
Vu19 40a	&	07:28:57.4	&	+69:13:29	&	2.48	&	70	&		&		\mbox{H\,{\sc ii}} region \\
Vu19 41	&	07:28:57.5	&	+69:13:33	&	0.42	&	50	&		&	\mbox{H\,{\sc ii}} region	\\
Vu19 42	&	07:28:57.5	&	+69:13:36	&	1.06	&	60	&		&	\mbox{H\,{\sc ii}} region	\\
Vu19 43	&	07:28:57.7	&	+69:12:57	&	13.42	&	100	&	0.06	&	giant \mbox{H\,{\sc ii}} region	\\
\hline
\end{tabular}
\vskip.25cm
$^{*}${\footnotesize Reddening corrected (Schlafly and Finkbeiner 2011).}

$^{**}${\footnotesize One arcsec corresponds to 17 pc for an assumed distance to NGC 2366 of 3.44 Mpc.}

$^{\dag}${\footnotesize This diameter is given according to the size of the corresponding shell detected with \textit{HST} (see Section 3.1).}

{\footnotesize Chomiuk and Wilcots (2009) radio-SNR IDs: $^{a}$ N2366-07; $^{b}$ N2366-12; $^{c}$ N2366-15; $^{d}$ N2366-16.}

 }\vskip.25cm

\noindent
\parbox{\textwidth}{
{\bf Table 2.} Continued.
\vskip.25cm
\begin{tabular}{@{\extracolsep{-1.0mm}}l c c c c c l @{}}
\hline
Object & Right   & Decl. & H$\alpha$ flux$^{*}$ & Diameter$^{**}$ & \hbox{[S\,{\sc ii}]}/H$\alpha$  &   Comment \\
ID & ascension &  &  [erg s$^{-1}$ cm$^{-2}$] &  [pc] & ratio &    \\
& $\alpha _{\mathrm{J2000}}$ & $\delta _{\mathrm{J2000}}$ & $\times10^{-15}$ & &  & \\
\hline \hline
Vu19 44$^{e, f}$	&	07:28:57.7	&	+69:13:41	&	30.40	&	160	&	0.06	& giant	\mbox{H\,{\sc ii}} region	\\
Vu19 45	&	07:28:58.4	&	+69:13:55	&	0.78	&	80	&		&	\mbox{H\,{\sc ii}} region	\\
Vu19 46	&	07:28:59.1	&	+69:13:27	&	2.25	&	40	&		&	\mbox{H\,{\sc ii}} region	\\
Vu19 47	&	07:29:00.6	&	+69:13:02	&	0.78	&	70	&		&	\mbox{H\,{\sc ii}} region	\\
Vu19 48	&	07:29:00.8	&	+69:13:38	&	0.85	&	80	&		&	diffuse	\\
Vu19 49	&	07:29:00.5	&	+69:12:02	&	1.52	&	90	&		&	\mbox{H\,{\sc ii}} region	\\
Vu19 50	&	07:29:00.8	&	+69:14:01	&	1.62	&	140	&		&	diffuse	\\
Vu19 51	&	07:29:01.3	&	+69:12:58	&	1.50	&	50	&		& \mbox{H\,{\sc ii}} region	\\
Vu19 52	&	07:29:01.0	&	+69:13:30	&	14.32	&	130	&		& \mbox{H\,{\sc ii}} region		\\
Vu19 53	&	07:29:00.8	&	+69:11:54	&	1.46	&	80	&		&	\mbox{H\,{\sc ii}} region	\\
Vu19 54	&	07:29:02.1	&	+69:12:55	&	10.57	&	130	&	0.05	&	\mbox{H\,{\sc ii}} region	\\
Vu19 55	&	07:29:02.8	&	+69:12:51	&	2.37	&	80	&		& \mbox{H\,{\sc ii}} region	\hspace{2cm}	\\
Vu19 56	&	07:29:02.7	&	+69:12:31	&	1.82	&	50	&		&	\mbox{H\,{\sc ii}} region	\\
Vu19 57	&	07:29:03.2	&	+69:12:35	&	1.71	&	80	&		&	\mbox{H\,{\sc ii}} region	\\
Vu19 58	&	07:29:03.0	&	+69:12:41	&	1.99	&	70	&		&	\mbox{H\,{\sc ii}} region	\\
Vu19 58	&	07:29:02.7	&	+69:11:54	&	0.52	&	50	&		&	\mbox{H\,{\sc ii}} region	\\
Vu19 60	&	07:29:03.9	&	+69:12:44	&	1.89	&	80	&		&	\mbox{H\,{\sc ii}} region	\\
Vu19 61	&	07:29:03.5	&	+69:14:02	&	36.83	&	180	&	0.10	& diffuse		\\
Vu19 62	&	07:29:04.3	&	+69:13:04	&	55.61	&	150	&	0.06	&	\mbox{H\,{\sc ii}} region	\\
Vu19 63	&	07:29:04.5	&	+69:13:11	&	9.59	&	$170\times80$	&		&	\mbox{H\,{\sc ii}} region	\\
Vu19 64	&	07:29:04.2	&	+69:13:31	&	3.67	&	80	&		&	\mbox{H\,{\sc ii}} region	\\
Vu19 65	&	07:29:04.3	&	+69:13:22	&	1.50	&	90	&		&	\mbox{H\,{\sc ii}} region	\\
Vu19 66	&	07:29:05.4	&	+69:12:57	&	4.22	&	90	&		&	\mbox{H\,{\sc ii}} region	\\
Vu19 67	&	07:29:11.4	&	+69:15:55	&	15.66	&	520	&		&	giant superbubble	\\
Vu19 68	&	07:29:13.1	&	+69:15:59	&	2.61	&	100	&		&	\mbox{H\,{\sc ii}} region	\\
\hline
\end{tabular}
\vskip.25cm
$^{*}${\footnotesize  Reddening corrected (Schlafly and Finkbeiner 2011). }

$^{**}${\footnotesize One arcsec corresponds to 17 pc for an assumed distance to NGC 2366 of 3.44 Mpc.}

{\footnotesize Chomiuk and Wilcots (2009) radio-SNR IDs: $^{e}$ N2366-18.}

{\footnotesize $^{f}$Also designated as N2366BG14 background galaxy in the field of NGC 2366 (Drissen et al. 2000), and compact star cluster (Billett et al. 2002).}

 }\vskip.75cm

\begin{multicols}{2}

\section{4. SUMMARY}

In this paper, we presented optical observations of the nearby irregular galaxy NGC 2366. This was the first time that this galaxy was observed through \hbox{[S\,{\sc ii}]} filter, i.e.~with
the intention to detect optical SNR candidates. Beside 64 probable \hbox{H\,{\sc ii}} regions and filaments, and three superbubbles, we suggested two objects as SNR {candidates} - {Vu19 5 and Vu19 34}. We underline that our observations were done with rather poor seeing condition (up to 3\uu25=55 pc for FOV1 and 2\uu0=35 pc for FOV2) which probably led to non-detection of potential SNRs with  smaller than {seeing-limited} diameter,
and SNRs in crowded starforming regions. {Archival \textit{HST} WFPC2 image through F656N filter showed shell-like structure at the position of SNR {candidate} Vu19 5}. Also, archival {\it XMM-Newton} observations suggest possible faint X-ray counterparts to two of our optical SNR candidates, which have also been detected with VLA and previously designated as radio-SNRs by Chomiuk and Wilcots (2009). In addition, according to their optical appearance and former identifications as background galaxies, {resolved with \textit{HST}}, we discard two previous radio
SNR candidates in this galaxy. This suggests that, so far, NGC 2366 galaxy hosts three SNR candidates,{ two visible in optical, X-ray and radio-domain,}  and one radio SNR candidate.

\acknowledgements{This research has been supported by the Ministry of Education, Science {and Technological Development} of the Republic of Serbia through the project No.~176005
"Emission nebulae: structure and evolution" and it is a part of the joint project of Serbian Academy of Sciences and Arts and Bulgarian Academy of Sciences
"Optical search for supernova remnants and {\hbox{H\,{\sc ii}}} regions in nearby galaxies (NGC 2366 and NGC 5585)". Authors thank Dragana \'{C}iprijanovi\'{c} for help with image editing. Authors gratefully acknowledge observing grant
support from the Institute of Astronomy and Rozhen National Astronomical Observatory, Bulgarian Academy of Sciences. This research has made use of
data obtained from the 3XMM XMM-Newton serendipitous source catalogue compiled by the 10 institutes of the {\it XMM-Newton} Survey Science Centre selected by ESA.}

\references

Billett, O. H. and Hunter, D. A.: 2002,  \journal{Astron. J.}, \vol{123}, 1454.

Blair, W. P. and Long, K. S.,: 1997, \journal{Astrophys. J. Suppl. Series}, \vol{108}, 261.

Bozzetto, L. M., Filipovi\'{c}, M. D., Vukoti\'{c}, B., Pavlovi\'{c}, M. Z., Uro\v{s}evi\'{c}, D., Kavanagh, P. J., Arbutina, B., Maggi, P. et al.: 2017, \journal{Astrophys. J. Suppl. Series}, \vol{230}, 2.

Chomiuk, L. and Wilcots, E. M.: 2009, \journal{Astron. J.}, \vol{137}, 3869.

Crawford, E. J., Filipovi\'{c}, M. D., Haberl, F., Pietsch, W., Payne, J. L. and de Horta, A. Y.: 2010,  \journal{Astron. Astrophys.}, \vol{518}, 35.

Crawford, E. J., Filipovi\'{c}, M. D., McEntaffer, R. L., Brantseg, T., Heitritter, K., Roper, Q., Haberl, F. and Uro\v{s}evi\'{c}, D.: 2014, \journal{Astron. J.}, \vol{148}, 99.

de Vaucouleurs, G., de Vaucouleurs, A., Corwin, H. G., Jr., et al.: 1991, Third
Reference Catalogue of Bright Galaxies, Springer, New York.

Drissen, L.,  Jean-ReneRoy, J.-R., Robert, C. and  Devost, D.: 2000, \journal{Astron. J.}, \vol{119}, 688.

D'Odorico, S., Dopita,M. A. and Benvenuti, P.: 1980, \journal{Astron.  Astrophys. Suppl. Series}, \vol{40}, 67.

Fesen, R. A., Blair,W. P. and Kirshner, R. P.: 1985,  \journal{Astrophys. J.}, \vol{292}, 29.

Filipovi\'{c}, M. D., Haynes, R. F., White, G. L. and Jones, P. A.: 1998, \journal{Astron.  Astrophys. Suppl. Series}, \vol{130}, 421.

Hodge, P. W. and Kennicutt, R. C. Jr.: 1983, \journal{Astron. J.}, \vol{88}, 296.

Kavanagh, P. J., Vink, J., Sasaki, M., Chu, Y.-H., Filipovi\'{c}, M. D., Ohm, S., Haberl, F., Manojlovi\'{c}, P. and Maggi, P.: 2019, \journal{Astron.  Astrophys.}, \vol{621}, 138.

Long, K. S.: 2017, in "Handbook of Supernovae", eds. A. W. Alsabti and P. Murdin, Springer International Publishing.

Mathewson, D. S. and Clarke, J. N.: 1973, \journal{Astrophys. J.}, \vol{178}, 105.

Matonick,  D. M. and Fesen, R. A.: 1997, \journal{Astrophys. J. Suppl. Series}, \vol{112}, 49.

Maxted, N. I., Filipovi\'{c}, M. D., Sano, H., Allen, G. E., Pannuti, T. G., Rowell, G. P., Grech, A., Roper, Q. et al.: 2018, \journal{Astrophys. J.}, \vol{866}, 76.

M\'{e}ndez, D. I., Esteban, C., Filipovi\'{c}, M. D., Ehle, M., Haberl, F., Pietsch, W. and Haynes, R. F.: 1999,  \journal{Astron.  Astrophys.}, \vol{349}, 801.

Micheva, G., Oey, M.~S., Jaskot, A.~E. and James, B.~L.: 2017, \journal{Astrophys. J.}, \vol{845}, 165.

Oke, J. B.: 1990, \journal{Astron. J.}, \vol{99}, 1621.

Roy, J.-R, Belley, J., Dutil, Y. and Martin, P.: 1996, \journal{Astrophys. J.}, \vol{460}, 284.

Sabin, L., Parker, Q. A., Contreras, M. E., Olguín, L., Frew, D. J., Stupar, M., Vázquez, R., Wright, N. J. et al.: 2013,  \journal{Mon. Not. R. Astron. Soc.}, \vol{431}, 279.

Sano, H. et al.: 2017, \journal{Astrophys. J.}, \vol{843}, 61.

Sasaki, M., Haberl, F., Henze, M., Saeedi, S., Williams, B.~F., Plucinsky, P.~P., Hatzidimitriou, D., Karampelas, A. et al.: 2018, \journal{Astron. Astrophys.}, \vol{620}, 28.

Schlafly, E. F. and  Finkbeiner, D. P.: 2011, \journal{Astrophys. J.}, \vol{737}, 103.

Stupar, M., Parker, Q. A. and Filipovi\'{c}, M. D.: 2008, \journal{Mon. Not. R. Astron. Soc.}, \vol{390}, 1037.

Sturm, R., Haberl, F., Pietsch, W., Ballet, J., Hatzidimitriou, D., Buckley, D. A. H., Coe, M., Ehle, M. et al.: 2013, \journal{Astron.  Astrophys.}, \vol{558}, 3.

Thuan, T.~X., Bauer, F.~E. and Izotov, Y.~I.: 2014, \journal{Mon. Not. R. Astron. Soc.}, \vol{441}, 1841.

Tolstoy, E., Saha, A., Hoessel, J. G. and McQuade, K.: 1995, \journal{Astron. J.}, \vol{110}, 164.

van Eymeren, J., Bomans, D. J., Weis, K. and Dettmar, R.-J.: 2007, \journal{Astron.  Astrophys.}, \vol{474}, 67.

Vu\v{c}eti\'{c}, M. M., Arbutina, B., Uro\v{s}evi\'{c}, D., Dobard\v{z}i\'{c}, A., Pavlovi\'{c}, M. Z., Pannuti, T. G. and Petrov, N.: 2013, \journal{Serb. Astron. J.}, \vol{187}, 11.

Vu\v{c}eti\'{c}, M. M., Arbutina, B. and Uro\v{s}evi\'{c}, D.: 2015, \journal{Mon. Not. R. Astron. Soc.}, \vol{446}, 943.

Watson, M. G. et al.: 2009, \journal{Astron.  Astrophys.}, \vol{493}, 339.
	
Yin, J., Matteucci, F. and Vladilo, G.: 2011, \journal{Astron.  Astrophys.}, \vol{531}, 136.

\endreferences

\end{multicols}

\vfill\eject

{\ }



\naslov{OPTIQKA POSMATRANJA BLISKE GALAKSIJE $\mathbf{NGC 2366}$ KROZ USKOPOJASNE FILTERE $\mathbf{[SII]}$ I $\mathbf{H}\alpha$. STATUS OSTATAKA SUPERNOVIH}

\authors{M. M. Vu\v{c}eti\'{c}$^{1}$, D. Oni\'{c}$^{1}$, N. Petrov$^{2}$, A. \'{C}iprijanovi\'{c}$^{1}$ and M. Z. Pavlovi\'{c}$^{1}$}

\vskip3mm

\address{$^1$Department of Astronomy, Faculty of Mathematics,
University of Belgrade\break Studentski trg 16, 11000 Belgrade,
Serbia}

\Email{mandjelic}{math.rs, donic@math.rs, aleksandra@math.rs; marko@math.rs}

\address{$^{2}$Institute of Astronomy and National Astronomical Observatory, Bulgarian
Academy of Sciences, 72 Tsarigradsko Shosse Blvd, BG-1784 Sofia, Bulgaria}

\Email{nip.sob}{gmail.com}

\vskip.7cm


\centerline{UDK \udc}


\centerline{\rit Profesionalni rad}

\vskip.7cm

\begin{multicols}{2}
{


{\rrm U radu je prezentovana detekcija 67 $\mathrm{HII}$ regiona i dva optiqka kandidata za   ostatake supernovih (OSN) u obli\zz njoj nepravilnoj
galaksiji $\mathrm{NGC 2366}$. Detekcija je izvr\ss ena upotrebom kriterijuma vezanog za odnos $\mathrm{[SII]}$ i
$\mathrm{H}\alpha$ linija, koriste\cc i posmatranja sa dvometarskog teleskopa Nacionalne astronomske opservatorije Ro\zz
en u Bugarskoj. U radu su dati polo\zz aji, dijametri, kao i $\mathrm{H}\alpha$ i $\mathrm{[SII]}$ fluksevi detektovanih objekata u dva posmatrana vidna polja u galaksiji $\mathrm{NGC 2366}$. Na osnovu arhivskih posmatranja  rentgenskog teleskopa $\mathrm{XMM-Newton}$ uoqena su dva izvora slabog sjaja koja odgovaraju  optiqkim kandidatima za OSN. Takodje, na osnovu optiqkih posmatranja zakljuqujemo da je prethodno predlo\zz ena klasifikacija dva objekta kao radio-OSN verovatno pogre\ss na, te da su to najverovatnie pozadinske galaksije.}

}
\end{multicols}

\end{document}